\documentclass[twocolumn]{jpsj3}
\usepackage{graphicx}
\usepackage{dcolumn}
\usepackage{bm}
\usepackage{ulem}
\usepackage{braket}
\usepackage{amsmath}
\usepackage{color}

\title{
Antiferro skyrmion crystal phases in a synthetic bilayer antiferromagnet\\ under an in-plane magnetic field
}

\author{Satoru Hayami}
\inst{
Graduate School of Science, Hokkaido University, Sapporo 060-0810, Japan 
}

\abst{
We theoretically propose a generation of an antiferro skyrmion crystal (AF-SkX) in a synthetic bilayer antiferromagnetic system under an external magnetic field. 
By performing the simulated annealing for a spin model on a centrosymmetric bilayer triangular lattice, we find that the interplay among the layer-dependent Dzyaloshinskii-Moriya interaction, the easy-plane single-ion anisotropy, the interlayer exchange interaction, and the in-plane magnetic field leads to instability toward the AF-SkX. 
The obtained AF-SkX consists of the skyrmion layer and anti-skyrmion layer with opposite skyrmion numbers, which results in no skyrmion number as well as spin scalar chirality in the whole system. 
We also show that a ferri-type skyrmion crystal, where one of the layers has the skyrmion number of one and the other does not, appears when the magnitude of the magnetic field is varied. 
Our results provide a guideline to experimentally search for the AF-SkX in bulk and layered materials. 
}

\begin{document}
\maketitle

\section{Introduction}
Noncoplanar magnets, where the spins align neither in a line nor on a plane, have attracted growing interest in various fields of condensed matter physics, such as spintronics, topological magnets, and multiferroics~\cite{batista2016frustration}. 
When noncoplanar spin textures are coupled to itinerant electrons, unconventional quantum transport, such as the topological Hall effect and nonreciprocal transport, occurs through a fictitious magnetic field based on the Berry phase mechanism, which does not rely on relativistic spin-orbit coupling~\cite{Loss_PhysRevB.45.13544, Ye_PhysRevLett.83.3737, Haldane_PhysRevLett.93.206602, Nagaosa_RevModPhys.82.1539, Xiao_RevModPhys.82.1959,Ohgushi_PhysRevB.62.R6065,tatara2002chirality,Shindou_PhysRevLett.87.116801,Martin_PhysRevLett.101.156402,Hayami_PhysRevB.106.014420,Hayami_doi:10.7566/JPSJ.91.094704}. 
As such a fictitious field reaches as large as $10^3$--$10^4$~T, noncoplanar magnets may be useful to realize high-speed and low-power devices for spintronic applications~\cite{Zutic_RevModPhys.76.323,jungwirth2016antiferromagnetic,Baltz_RevModPhys.90.015005,vsmejkal2018topological,jungfleisch2018perspectives}. 

One of the most typical examples to have noncoplanar spin textures is a magnetic skyrmion, which is characterized by an integer topological number (skyrmion number)~\cite{skyrme1962unified, Bogdanov89, Bogdanov94,nagaosa2013topological}. 
Once the skyrmion is periodically aligned under the lattice structures, which is so-called the skyrmion crystal (SkX), a large magnetoelectric response triggered by the Berry phase is expected~\cite{Bruno_PhysRevLett.93.096806, Neubauer_PhysRevLett.102.186602, Kanazawa_PhysRevLett.106.156603}. 
In spite of complicated spin textures in the SkX, it has been observed in various materials under both noncentrosymmetric~\cite{Muhlbauer_2009skyrmion,yu2010real,yu2011near,seki2012observation, Adams2012, Seki_PhysRevB.85.220406,nayak2017discovery, Kurumaji_PhysRevLett.119.237201} and centrosymmetric~\cite{kurumaji2019skyrmion,hirschberger2019skyrmion, Hirschberger_10.1088/1367-2630/abdef9,khanh2020nanometric, Yasui2020imaging,khanh2022zoology,takagi2022square} lattice structures~\cite{Tokura_doi:10.1021/acs.chemrev.0c00297}. 
Simultaneously, theoretical studies have revealed the stabilization mechanisms of the SkX: the Dzyaloshinskii-Moriya (DM) interaction~\cite{dzyaloshinsky1958thermodynamic,moriya1960anisotropic,rossler2006spontaneous,Yi_PhysRevB.80.054416,Hayami_PhysRevB.105.224423}, dipolar interaction~\cite{Utesov_PhysRevB.103.064414,Utesov_PhysRevB.105.054435,tong2018skyrmions}, frustrated exchange interaction~\cite{Okubo_PhysRevLett.108.017206,leonov2015multiply,Lin_PhysRevB.93.064430,Hayami_PhysRevB.93.184413,Hayami_PhysRevB.94.174420,Lin_PhysRevLett.120.077202,hayami2022skyrmion,Hayami_PhysRevB.105.174437,Okigami_doi:10.7566/JPSJ.91.103701}, long-range spin interaction~\cite{Ruderman,Kasuya,Yosida1957,Wang_PhysRevLett.124.207201,Mitsumoto_PhysRevB.104.184432,Mitsumoto_PhysRevB.105.094427,Kobayashi_PhysRevB.106.L140406}, multiple spin interaction~\cite{heinze2011spontaneous,Ozawa_PhysRevLett.118.147205,Hayami_PhysRevB.95.224424,Christensen_PhysRevX.8.041022,Hayami_PhysRevB.99.094420,hayami2020multiple,Eto_PhysRevB.104.104425,Nikoli_PhysRevB.103.155151,hayami2021topological,Hayami_10.1088/1367-2630/ac3683,hayami2021phase,wang2021skyrmion,Eto_PhysRevLett.129.017201,hayami2022multiple,hayami2022widely}, anisotropic exchange interaction~\cite{Hayami_PhysRevLett.121.137202,amoroso2020spontaneous,Hayami_PhysRevB.103.024439,Hayami_PhysRevB.103.054422,yambe2021skyrmion,Wang_PhysRevB.103.104408,amoroso2021tuning,Hayami_PhysRevB.105.104428,Kato_PhysRevB.105.174413,Yambe_PhysRevB.106.174437}, and their combination~\cite{Kathyat_PhysRevB.103.035111,Hayami_PhysRevB.104.094425,hayami2021field, Hayami_10.1088/2515-7639/acab89}.
Thus, many potential situations exist so as to engineer and design the SkX irrespective of the lattice structures and microscopic mechanisms.  

On the other hand, an SkX in synthetic antiferromagnets, which is characterized by an antiferroic alignment of the magnetic skyrmions under multi-sublattice structures, has been still rare compared to the SkX~\cite{Bogdanov_PhysRevB.66.214410,gobel2021beyond,yambe2022ferrochiral}. 
We here refer to such an SkX as antiferro SkX (AF-SkX), since the AF-SkX consists of the skyrmions with an opposite skyrmion number in different sublattices as a consequence of the staggered alignment of spins like $\bm{S}^{(\rm A)}=-\bm{S}^{(\rm B)}$ ($\bm{S}^{(\eta)}$ is the spins for sublattice $\eta$). In such a situation, the emergent magnetic field is canceled out in the whole system; the topological Hall effect vanishes~\cite{Gobel_PhysRevB.96.060406}. 
Meanwhile, the spin textures in the AF-SkX are topologically protected, which results in other topological phenomena, such as the topological spin Hall effect~\cite{buhl2017topological, Gobel_PhysRevB.96.060406, Akosa_PhysRevLett.121.097204}. 
Moreover, it can be promising for spintronic applications owing to the absence of the skyrmion Hall effect; the isolated AFM skyrmion can move parallel to an applied current without deviating~\cite{zhang2016antiferromagnetic, Barker_PhysRevLett.116.147203,zhang2016magnetic,jin2016dynamics, Shen_PhysRevB.98.134448,shen2019spin, Jin_PhysRevB.102.054419,Shen_PhysRevApplied.12.064033}. 
Thus, it is desired to find out the stabilization mechanism of the AF-SkX, which is one of the challenges in this active research area. 
However, such a condition especially for the case in the presence of an external magnetic field has not been clarified yet, since the magnetic field usually breaks the relationship of $\bm{S}^{(\rm A)}=-\bm{S}^{(\rm B)}$ in the spin textures. 

In the present study, we theoretically investigate the realization of the AF-SkX in centrosymmetric layered magnets under the external magnetic field. 
The results are obtained by performing the simulated annealing for an effective spin model with the intralayer momentum-resolved interaction and interlayer exchange interaction on a bilayer triangular lattice. 
By taking into account the effect of the staggered DM interaction, easy-plane anisotropy, and in-plane magnetic field, we show that the AF-SkX is robustly stabilized in the ground state. 
The AF-SkX exhibits the opposite skyrmion number for different layers so that the spin scalar chirality in the whole system vanishes. 
This AF-SkX is qualitatively different from the previous antiferromagnetic SkX stabilized at zero field~\cite{zhang2016antiferromagnetic, Barker_PhysRevLett.116.147203, Bessarab_PhysRevB.99.140411, Kravchuk_PhysRevB.99.184429,Zarzuela_PhysRevB.100.100408,legrand2020room} and sublattice-dependent SkX with a uniform scalar chirality~\cite{Rosales_PhysRevB.92.214439, Diaz_hysRevLett.122.187203, Osorio_PhysRevB.96.024404, LIU201925,gao2020fractional, liu2020theoretical, Tome_PhysRevB.103.L020403, mukherjee2021antiferromagnetic, Mukherjee_PhysRevB.103.134424, Rosales_PhysRevB.105.224402}.
We also find that two ferri-type SkX (Ferri-SkX) phases, which consist of the skyrmion layer and topologically-trivial magnetic layer, appear in the vicinity of the AF-SkX phase. 
Our results provide a way of realizing the AF-SkX by applying the magnetic field far from zero magnetic field, which will stimulate further experimental observations of exotic topological spin crystals beyond the conventional SkX.

The remainder of this paper is organized as follows. 
In Sec.~\ref{sec: Model and method}, we introduce the spin model on the bilayer triangular lattice. 
We also outline simulated annealing to obtain the ground-state spin configuration. 
Then, we show the magnetic phase diagram while changing the interlayer exchange interaction and the in-plane magnetic field in Sec.~\ref{sec: Results}. 
We discuss the spin and scalar chirality configurations in the AF-SkX as well as the other multiple-$Q$ and single-$Q$ states. 
Finally, we summarize the paper in Sec.~\ref{sec: Summary}. 

\section{Model and method}
\label{sec: Model and method}

\begin{figure}[t!]
\begin{center}
\includegraphics[width=1.0 \hsize ]{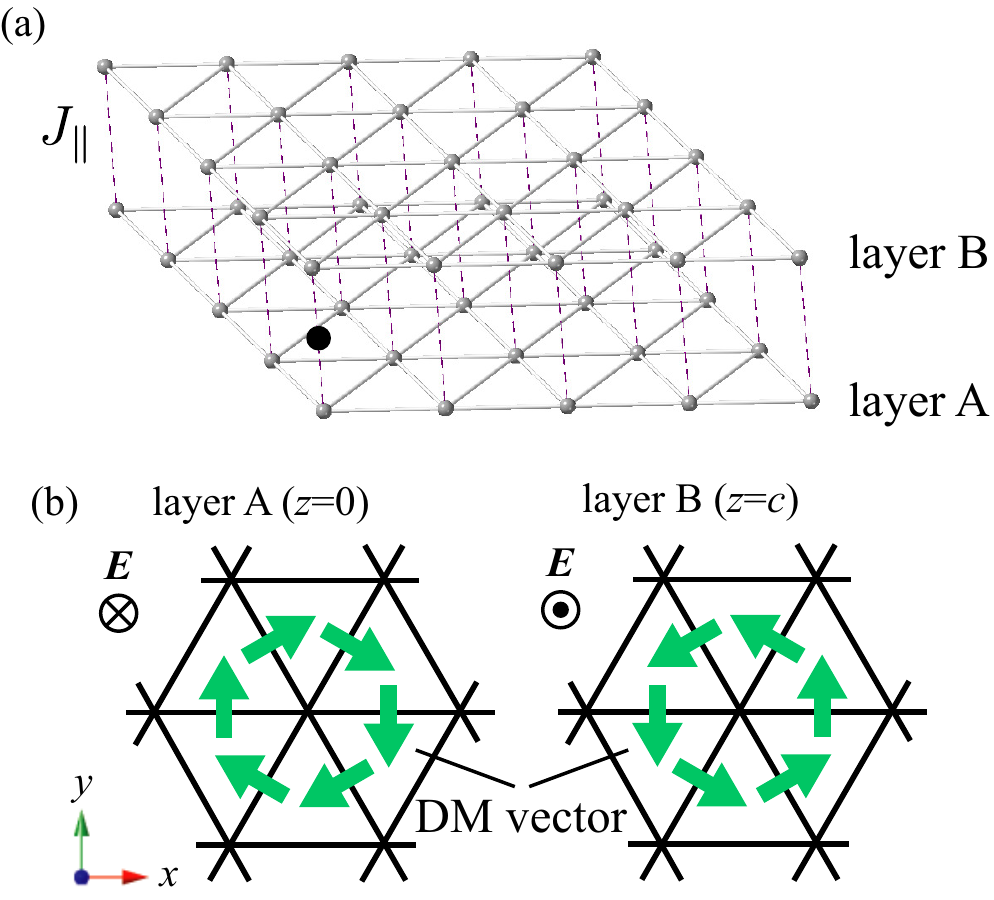} 
\caption{
\label{fig: lattice}
(Color online) 
(a) Bilayer triangular lattice consisting of layer A and layer B; $J_{\parallel}$ represents the interlayer exchange coupling. 
The black circle denotes the inversion center. 
(b) The layer-dependent DM interaction for (left) layer A and (right) layer B in the presence of the local crystalline electric field $\bm{E}$. 
The green arrows stand for the DM vectors, which are opposite for two layers. 
}
\end{center}
\end{figure}

We investigate the multiple-$Q$ instability in the bilayer triangular-lattice system. 
The bilayer structure consists of two triangular layers on the $xy$ plane, which are denoted as layer A and layer B, as shown in Fig.~\ref{fig: lattice}(a).  
The two layers are separated along the $z$ axis by $c=1$, where the lattice sites are located at the same $xy$ position; the lattice constant of the triangular lattice is set as unity. 
The effect of the difference along the $z$ and inplane directions is taken into account as the different magnitudes of the interactions [see $J$ and $J_{\parallel}$ in Eqs.~(\ref{eq:Ham_perp}) and (\ref{eq:Ham_parallel})].
It is noted that an inversion center is present at the nearest-neighbor bond between layers A and B denoted as the black circle in Fig.~\ref{fig: lattice}(a), although there is no local inversion center in each layer. 
Specifically, we consider the situation where the total bilayer system belongs to the $D_{\rm 6h}$ point group, and the individual layer has the $C_{\rm 6v}$ symmetry; the local crystalline electric field $\bm{E}$ with the same magnitude but the opposite direction is present in the $z$ direction [Fig.~\ref{fig: lattice}(b)], which is a source of the layer-dependent DM interaction, as described below. 
In such a situation, the DM vector lies in the $xy$ plane like the Rashba-induced case~\cite{Kim_PhysRevLett.111.216601,Kundu_PhysRevB.92.094434,hayami2016emergent}.

In such a layered lattice structure without local inversion symmetry, we consider the spin Hamiltonian $\mathcal{H}$, which is given by 
\begin{align}
\label{eq:Ham}
\mathcal{H}=&\sum_{\eta}\mathcal{H}^{\perp}_{\eta}+\mathcal{H}^{\parallel}+\mathcal{H}^{{\rm loc}}, 
\end{align}
where $\mathcal{H}^{\perp}$, $\mathcal{H}^{\parallel}$, and $\mathcal{H}^{{\rm loc}}$ are explicitly represented as 
\begin{align}
\label{eq:Ham_perp}
\mathcal{H}^{\perp}_\eta=&  \sum_{\nu}\Big[ -J \bm{S}^{(\eta)}_{{\bm{Q}_\nu}} \cdot \bm{S}^{(\eta)}_{-\bm{Q}_{\nu}}- i   \bm{D}^{(\eta)}_{\bm{Q}_\nu} \cdot ( \bm{S}^{(\eta)}_{\bm{Q}_{\nu}} \times \bm{S}^{(\eta)}_{-\bm{Q}_{\nu}}) \Big], \\
\label{eq:Ham_parallel}
\mathcal{H}^{\parallel}=& J_{\parallel} \sum_{\langle i,j \rangle} \bm{S}_i \cdot \bm{S}_{j},\\
\label{eq:Ham_Zeeman}
\mathcal{H}^{{\rm loc}}=&A^{\rm ion} \sum_{i} (S_i^z)^2-H \sum_i S_i^x.  
\end{align}
Here, $\bm{S}_i$ is the classical localized spin at site $i$ with the magnitude of $|\bm{S}_i|=1$. 
The intralayer Hamiltonian $\mathcal{H}^{\perp}$ in Eq.~(\ref{eq:Ham_perp}) consists of the Heisenberg-type isotropic exchange interaction $J$ in the first term and the antisymmetric DM interaction $\bm{D}_{\bm{Q}_\nu}^{(\eta)}$ in the second term. 
Both interaction terms are represented in momentum space; $\bm{S}^{(\eta)}_{\bm{Q}_{\nu}}$ with layer $\eta$ and wave vector $\bm{Q}_\nu$ is the Fourier transform of $\bm{S}_i$. 

To focus on the multiple-$Q$ states that arise from a superposition of finite-$Q$ spiral states, we consider the $\pm \bm{Q}_1=(Q,0)$, $\pm \bm{Q}_2=(-Q/2,\sqrt{3}Q/2)$, and $\pm \bm{Q}_3=(-Q/2,-\sqrt{3}Q/2)$ components of the interaction, which are related by threefold rotational symmetry of the triangular lattice~\cite{comment_higherorderinteraction}. 
We set $Q=\pi/3$ without loss of generality.  
Such a finite-$Q$ instability is obtained by considering the effect of magnetic frustration arising from the competing exchange interaction in real and/or momentum space. 
$J$ is independent of $\bm{Q}_\nu$, while $\bm{D}_{\bm{Q}_\nu}^{(\eta)}$ depends on $\bm{Q}_\nu$. 
We assume that the position of $\bm{Q}_\nu$ is fixed against the magnetic field for simplicity.
Owing to the polar-type crystalline electric field along the $z$ direction, the DM vector is perpendicular to both the intralayer nearest-neighbor bond direction and the $z$ direction, as shown in Fig.~\ref{fig: lattice}(b); the magnitude of the DM interaction is set as $|\bm{D}_{\bm{Q}_\nu}^{(\eta)}|=D$. 
In addition, the sign of the DM interaction is opposite for layers A and B due to the staggered crystalline electric field. 
Thus, $\bm{D}_{\bm{Q}_1}^{(\rm A)}=-\bm{D}_{\bm{Q}_1}^{(\rm B)}=(0, -D, 0)$, $\bm{D}_{\bm{Q}_2}^{(\rm A)}=-\bm{D}_{\bm{Q}_2}^{(\rm B)}=(\sqrt{3}D/2, D/2, 0)$, and $\bm{D}_{\bm{Q}_3}^{(\rm A)}=-\bm{D}_{\bm{Q}_3}^{(\rm B)}=(-\sqrt{3}D/2, D/2, 0)$. 
It is noted that $D$ can take an arbitrary value depending on the real-space interactions; we treat $D$ as the model parameter.

The interlayer Hamiltonian $\mathcal{H}^{\parallel}$ in Eq.~(\ref{eq:Ham_parallel}) includes the antiferromagnetic nearest-neighbor interaction between layers A and B, $J_{\parallel}>0$; $\langle i,j \rangle$ represents the nearest-neighbor pair between layers A and B. 
The local Hamiltonian $\mathcal{H}^{{\rm loc}}$ in Eq.~(\ref{eq:Ham_Zeeman}) consists of the easy-plane single-ion anisotropy $A^{\rm ion}>0$ and the in-plane magnetic field along the $x$ direction with the magnitude of $H$. 
It is noted that the in-plane magnetic field breaks the threefold rotational symmetry of the lattice structure, which makes $\bm{Q}_1$ and $\bm{Q}_{2,3}$ inequivalent. 
The recent theoretical studies have revealed that the interplay between a layer-dependent DM interaction and interlayer exchange interaction stabilizes the SkX with the net scalar chirality rather than the AF-SkX without the net scalar chirality under the out-of-plane magnetic field irrespective of the sign of the interlayer exchange interaction~\cite{Hayami_PhysRevB.105.014408, Hayami_PhysRevB.105.184426,hayami2022square,lin2021skyrmion}.

The model in Eq.~(\ref{eq:Ham}) has five independent parameters: $J$, $D$, $J_{\parallel}$, $A^{\rm ion}$, and $H$. 
Among them, we take $J=1$ as the energy unit of the model. 
Besides, we fix $D=0.05$ and $A^{\rm ion}=0.2$ so that the SkX is stabilized under the in-plane magnetic field in the single-layer limit ($J_{\parallel}=0$); the SkX tends to be destabilized without $A^{\rm ion}$ or easy-axis anisotropy $A^{\rm ion}<0$. Especially, only the single-$Q$ conical spiral state appears for $A^{\rm ion}=D=0$. 
We discuss the stability of the AF-SkX when $J_{\parallel}$ and $H$ are varied. 
For the ferromagnetic interlayer exchange interaction, i.e., $J_{\parallel}<0$, a ferro-type SkX with the same skyrmion number in both layers is stabilized as found in previous studies~\cite{Hayami_PhysRevB.105.014408, Hayami_PhysRevB.105.184426,hayami2022square,lin2021skyrmion}.

We construct the magnetic phase diagram of the model in Eq.~(\ref{eq:Ham}) at low temperatures by performing simulated annealing. 
The simulations are carried out with the standard Metropolis local updates for real-space spins, where the direction of spins is randomly chosen on the sphere with $|\bm{S}_i|=1$. 
In each simulation, we gradually reduce the temperature with a rate $T_{n+1}=\alpha T_n$, where $T_n$ is the $n$th-step temperature, to find the low-energy spin configurations.
We set the initial temperature $T_0/J=1$--10, final temperature $T_f/J=0.001$, and $\alpha=0.999995$. 
The initial spin configuration is randomly chosen.
At the final temperature, $10^5$--$10^6$ Monte Carlo sweeps are performed for measurements. 
We also start the simulations from the spin configurations obtained by the above procedure in the vicinity of the phase boundaries.
The total number of spins is taken as $N=2\times L^2=2\times 24^2$ under the periodic boundary conditions, where $L$ is the length of the triangular-lattice system along the $\bm{a}_1=(1,0)$ and $\bm{a}_2=(-1/2,\sqrt{3}/2)$ directions ($\bm{a}_1$ and $\bm{a}_2$ are the primitive lattice vectors). 

We calculate the spin- and chirality-related quantities to identify the magnetic phases. 
In the spin sector, we calculate the $\alpha=x,y,z$ component of the spin structure factor $S_{s\eta}^\alpha(\bm{q})$, which is given by 
\begin{align}
S_{s\eta}^\alpha(\bm{q})= \frac{1}{L^2} \sum_{i,j \in \eta} S^{\alpha}_i S^{\alpha}_j e^{i\bm{q}\cdot (\bm{r}_i-\bm{r}_j)}, 
\end{align}
where the site indices $i$ and $j$ are taken for the spins on layers $\eta={\rm A}$ and B. 
The magnetic moments at wave vector $\bm{q}$ and layer $\eta$ are given by $m^{\alpha}_{\eta\bm{q}}=\sqrt{S_{s\eta}^\alpha(\bm{q})/L^2}$. 
We also calculate $m^{xy}_{\eta\bm{q}}=\sqrt{(m^{x}_{\eta\bm{q}})^2+(m^{y}_{\eta\bm{q}})^2}$. 
The uniform magnetization per layer is given by $M^\alpha_{\eta}=(1/L^2)\sum_{i \in \eta}S^{\alpha}_{i}$. 

In the scalar chirality sector, we compute the uniform spin scalar chirality for layer $\eta$, which is represented by 
\begin{align}
\chi^{\rm sc}_{\eta} = \frac{1}{L^2} \sum_{\bm{R}\in \eta} \bm{S}_{i} \cdot (\bm{S}_j \times \bm{S}_k),
\end{align}
where $\bm{R}$ represents the position vector at the centers of triangles; the sites $i$, $j$, and $k$ form the triangle at $\bm{R}$ in the counterclockwise order. 
It is noted that there are upward and downward triangle plaquettes. 
The local scalar chirality is given by $\chi_{\bm{R}}=\bm{S}_{i} \cdot (\bm{S}_j \times \bm{S}_k)$.

\section{Results}
\label{sec: Results}

\begin{figure}[t!]
\begin{center}
\includegraphics[width=1.0 \hsize ]{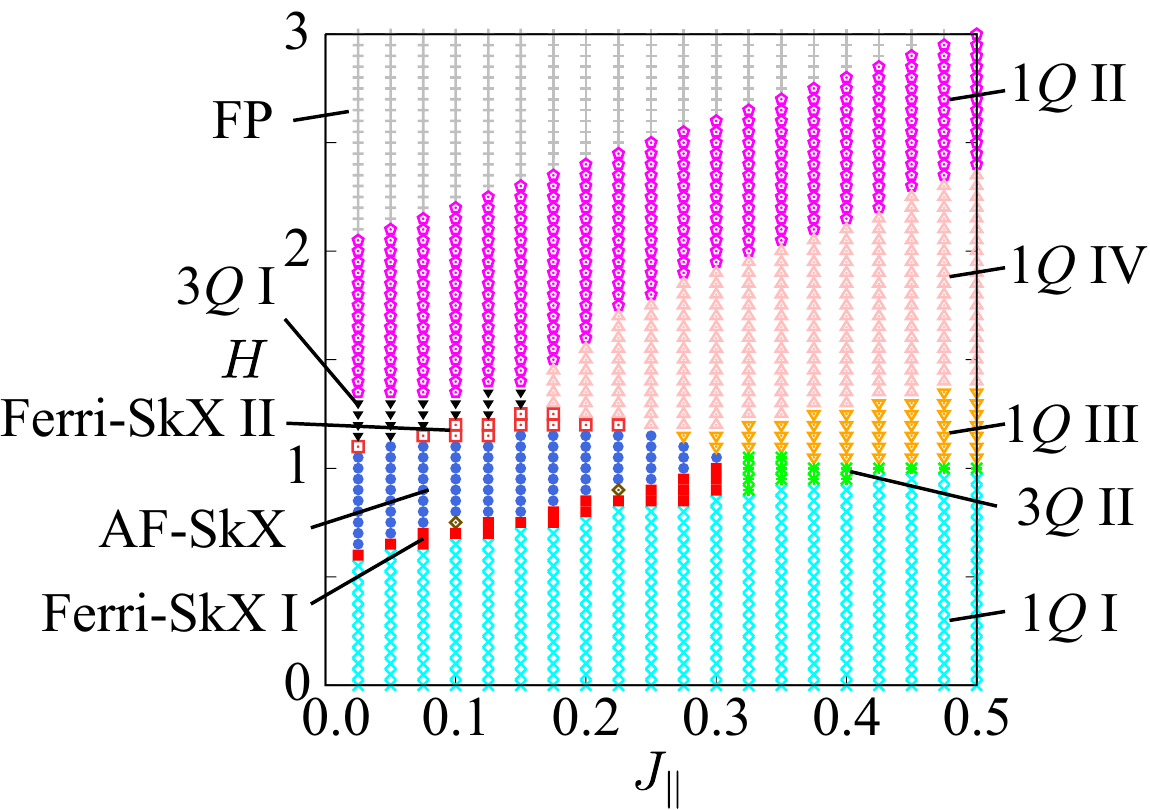} 
\caption{
\label{fig: PD}
(Color online) 
Magnetic phase diagram in the plane of the interlayer exchange interaction, $J_{\parallel}$, and the magnetic field along the $x$ direction, $H$, obtained by the simulated annealing at $T/J=0.001$.  
1$Q$ and 3$Q$ denote the single-$Q$ and triple-$Q$ states, respectively. 
Ferri-SkX, AF-SkX, and FP stand for the ferromagnetic skyrmion crystal, the antiferroic skyrmion crystal, and the fully-polarized state, respectively. 
The rhombus symbols at $(J_{\parallel}, H)=(0.1,0.75)$ and  $(J_{\parallel}, H)=(0.225,0.9)$ denote the AF-SkX' phase, which is characterized by the opposite skyrmion number for two layers like the AF-SkX but $\chi^{\rm sc}_{\rm A} \neq -\chi^{\rm sc}_{\rm B}$. 
}
\end{center}
\end{figure}

\begin{figure*}[t!]
\begin{center}
\includegraphics[width=1.0 \hsize ]{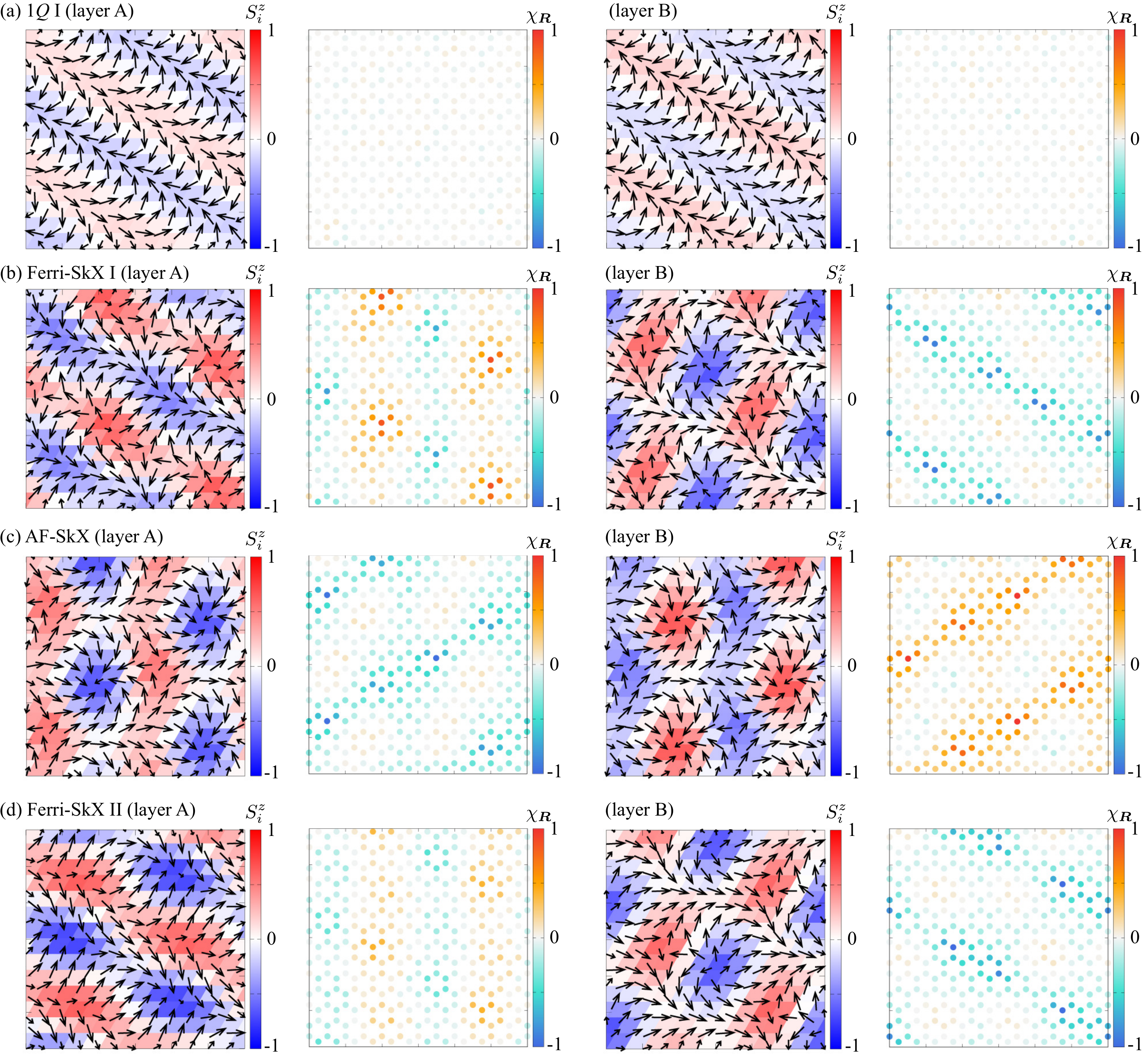} 
\caption{
\label{fig: spin}
(Color online) 
Left: Real-space spin configurations on layer A of (a) the $1Q$ I state at $J_{\parallel}=0.1$ and $H=0.5$, (b) the Ferri-SkX I at $J_{\parallel}=0.1$ and $H=0.7$, (c) the AF-SkX at $J_{\parallel}=0.1$ and $H=0.9$, and (d) the Ferri-SkX II at $J_{\parallel}=0.1$ and $H=1.2$. 
The arrows and colors represent the in-plane and out-of-plane components of the spin moment, respectively~\cite{comment_contour2}. 
Middle left: Real-space distribution of the spin scalar chirality at each triangle plaquette. 
Middle right and right: The spin and scalar chirality configurations on layer B corresponding to the left and middle left panels. 
}
\end{center}
\end{figure*}

\begin{figure*}[t!]
\begin{center}
\includegraphics[width=1.0 \hsize ]{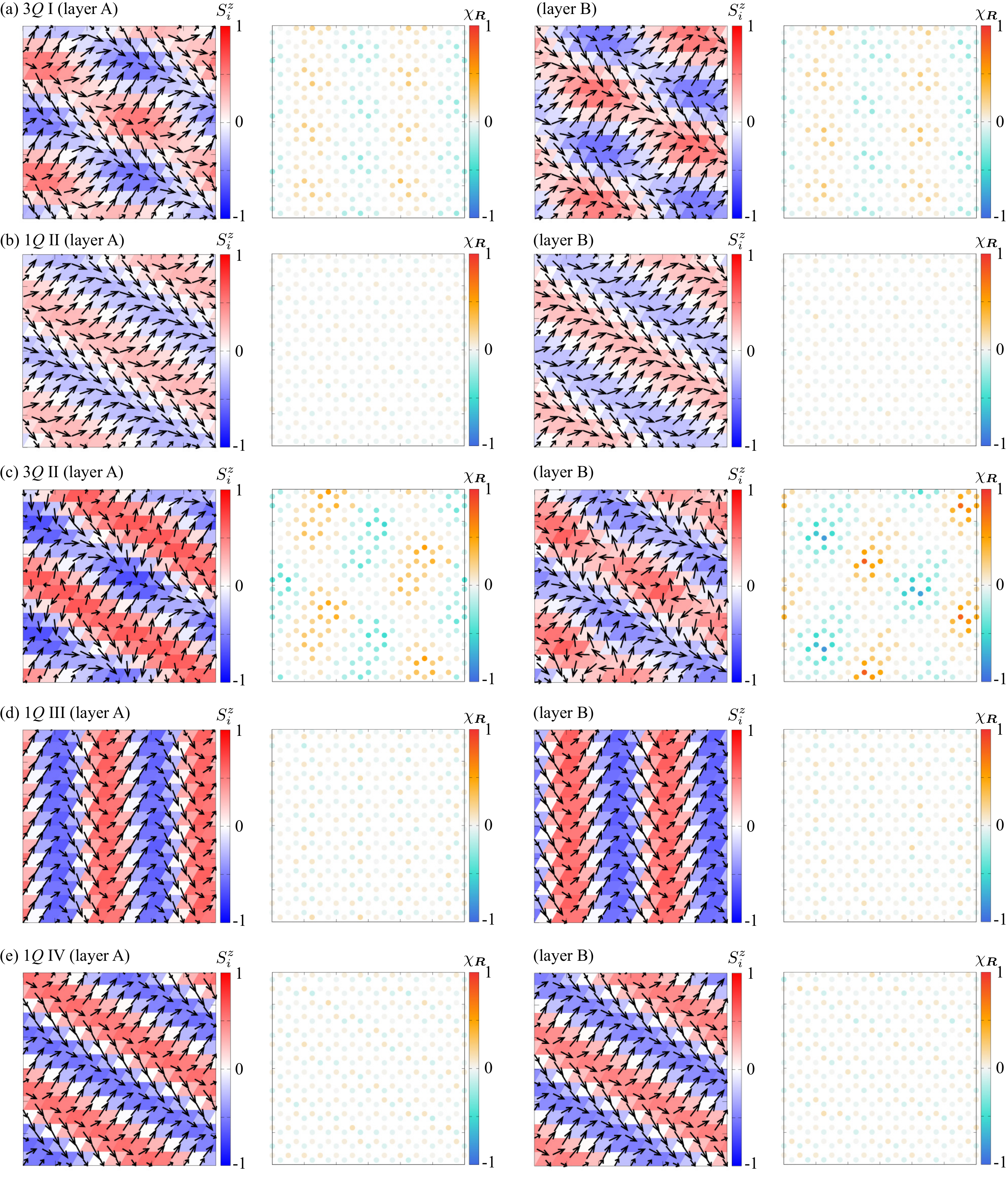} 
\caption{
\label{fig: spin2}
(Color online) 
The same plots as in Fig.~\ref{fig: spin} for (a) the $3Q$ I state at $J_{\parallel}=0.1$ and $H=1.3$, (b) the 1$Q$ II state at $J_{\parallel}=0.1$ and $H=1.5$, (c) the 3$Q$ II state at $J_{\parallel}=0.4$ and $H=1$, (d) the 1$Q$ III state at $J_{\parallel}=0.4$ and $H=1.2$, and (e) the 1$Q$ IV state at $J_{\parallel}=0.4$ and $H=1.5$. 
}
\end{center}
\end{figure*}

\begin{figure*}[t!]
\begin{center}
\includegraphics[width=1.0 \hsize ]{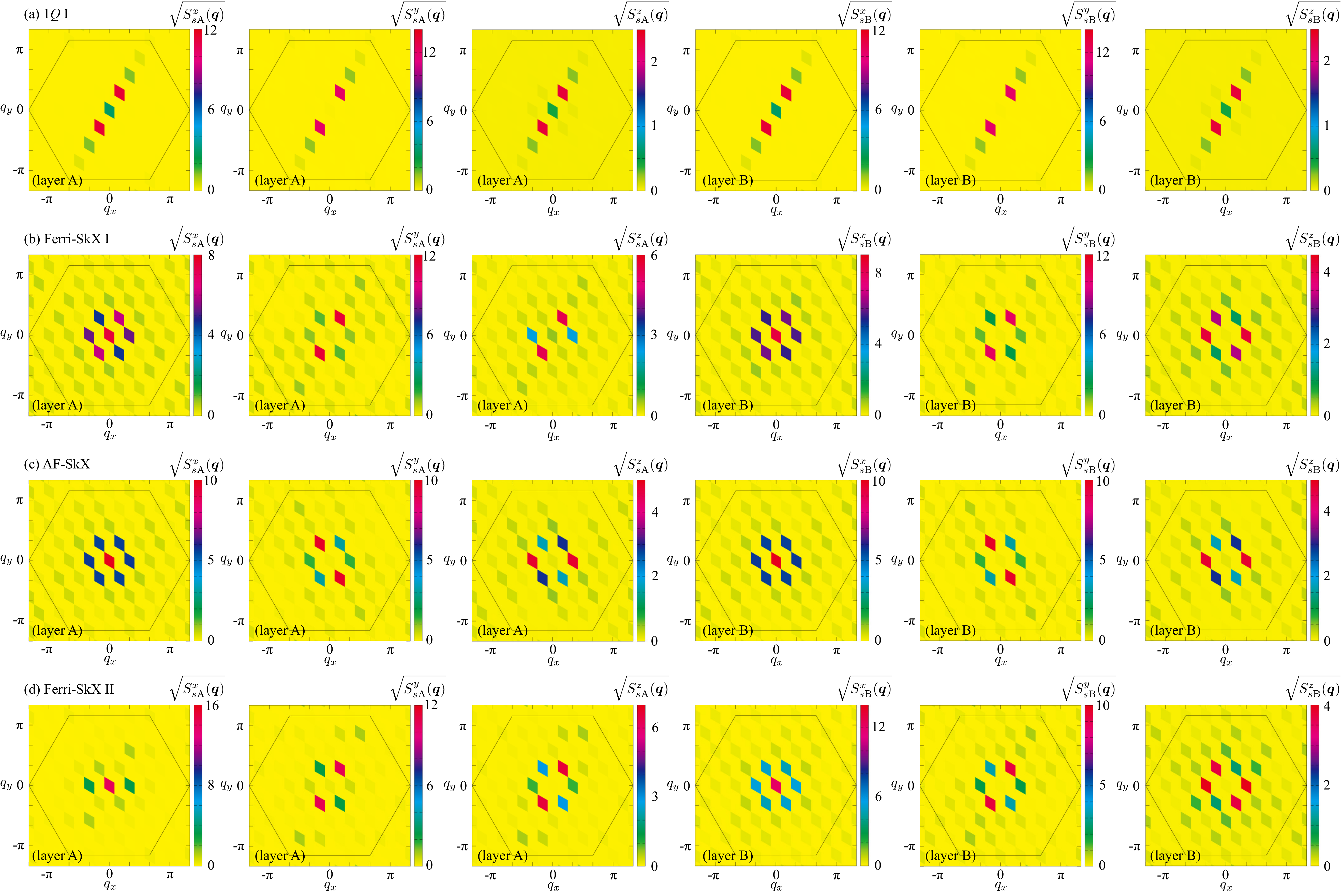} 
\caption{
\label{fig: Sq}
(Color online) 
Square root of the spin structure factor in (left) the $x$, (second left) $y$, and (third left) $z$ components for layer A. 
The right three panels represent the data for layer B corresponding to the left three ones. 
The data are (a) the $1Q$ I state at $J_{\parallel}=0.1$ and $H=0.5$, (b) the Ferri-SkX I at $J_{\parallel}=0.1$ and $H=0.7$, (c) the AF-SkX at $J_{\parallel}=0.1$ and $H=0.9$, and (d) the Ferri-SkX II at $J_{\parallel}=0.1$ and $H=1.2$. 
The black hexagons represent the first Brillouin zone. 
}
\end{center}
\end{figure*}

\begin{figure*}[t!]
\begin{center}
\includegraphics[width=1.0 \hsize ]{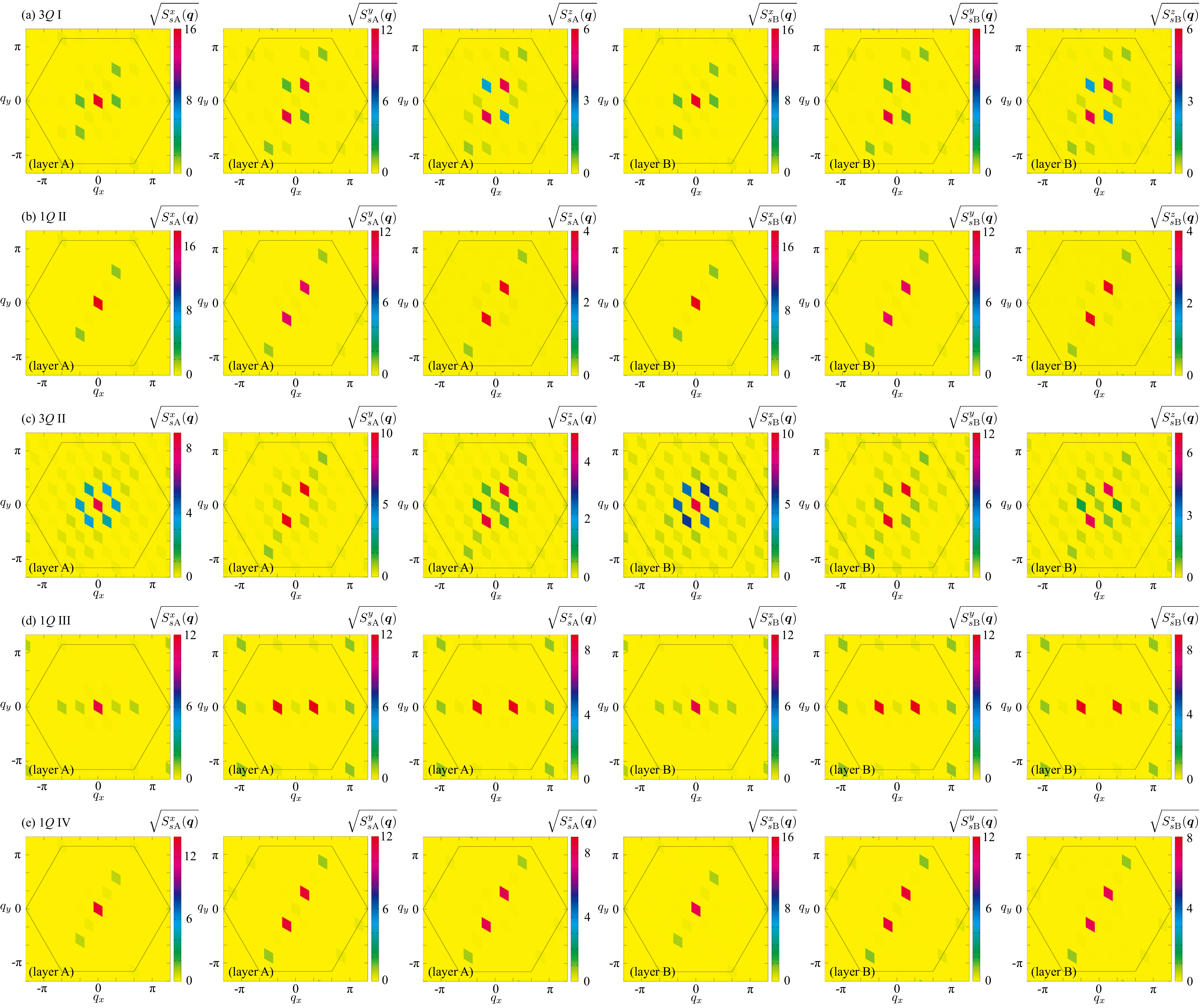} 
\caption{
\label{fig: Sq2}
(Color online) 
The same plots as in Fig.~\ref{fig: Sq} for (a) the $3Q$ I state at $J_{\parallel}=0.1$ and $H=1.3$, (b) the 1$Q$ II state at $J_{\parallel}=0.1$ and $H=1.5$, (c) the 3$Q$ II state at $J_{\parallel}=0.4$ and $H=1$, (d) the 1$Q$ III state at $J_{\parallel}=0.4$ and $H=1.2$, and (e) the 1$Q$ IV state at $J_{\parallel}=0.4$ and $H=1.5$. 
}
\end{center}
\end{figure*}

\begin{figure}[t!]
\begin{center}
\includegraphics[width=0.8 \hsize ]{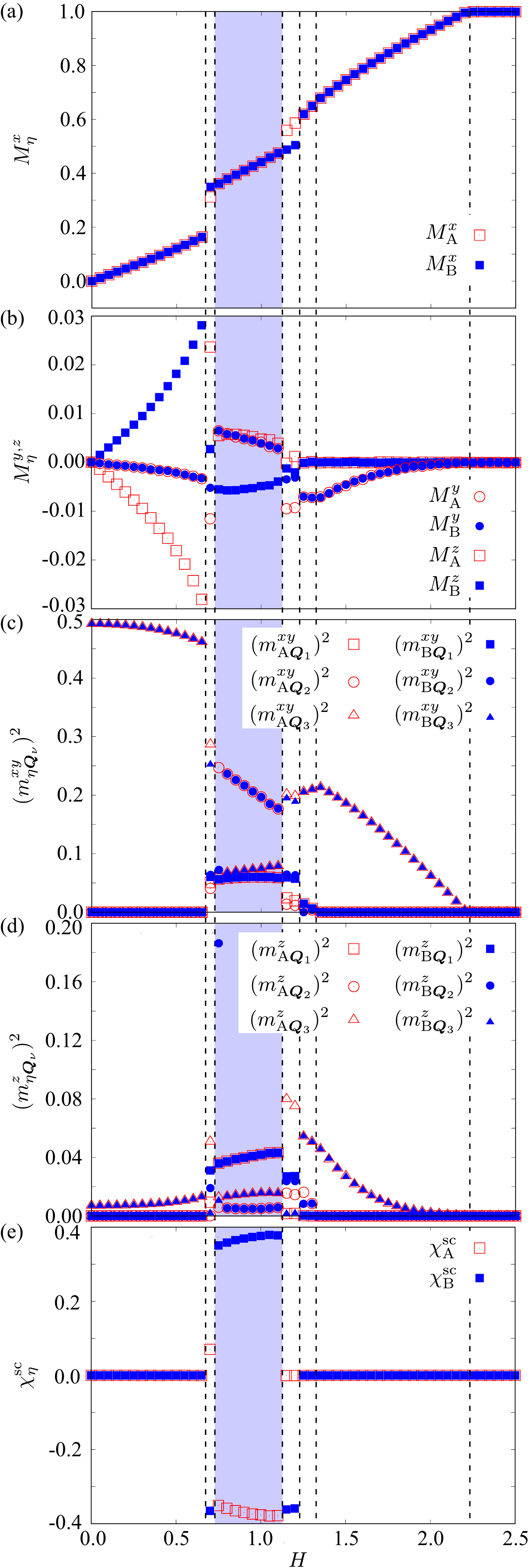} 
\caption{
\label{fig: mq_0.1}
(Color online) 
$H$ dependence of (a) $M^x_\eta$, (b) $M^{y,z}_\eta$, (c) $(m^{xy}_{\eta \bm{Q}_\nu})^2$, (d) $(m^{z}_{\eta \bm{Q}_\nu})^2$, and (e) $\chi^{\rm sc}_\eta$ for $\eta=$A and B at $J_{\parallel}=0.1$. 
The vertical dashed lines represent the phase boundary. 
The region drawn in blue represents the AF-SkX. 
}
\end{center}
\end{figure}

\begin{figure}[t!]
\begin{center}
\includegraphics[width=0.85 \hsize ]{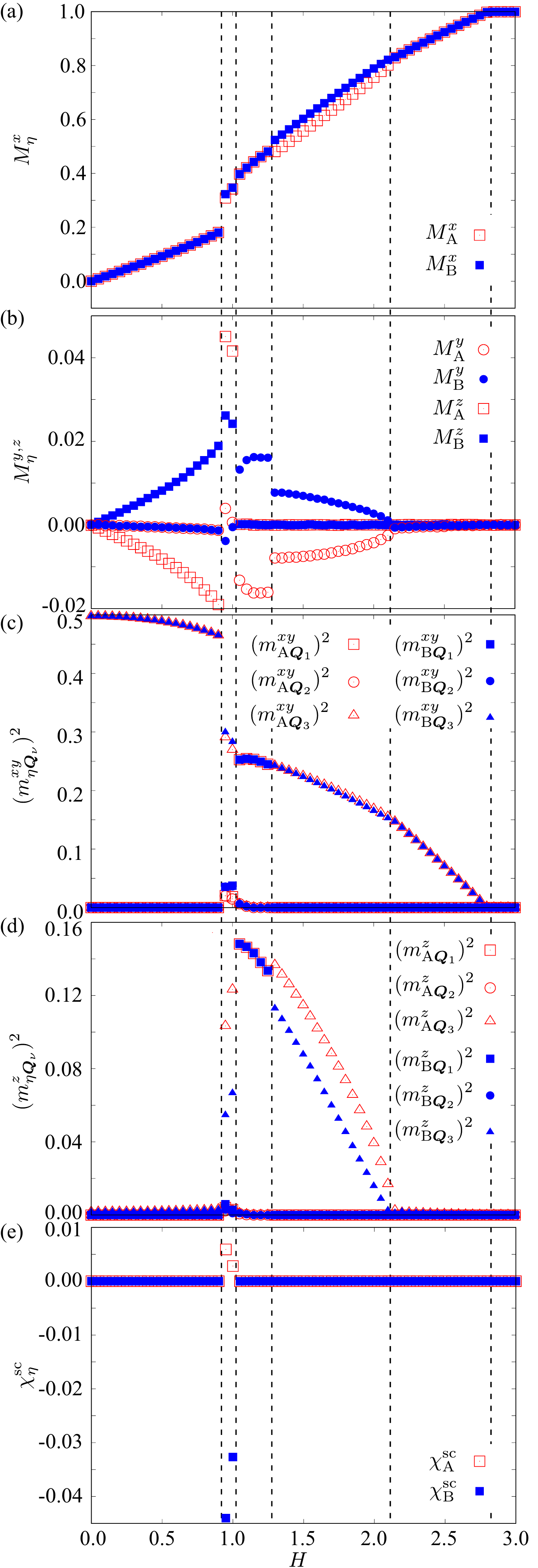} 
\caption{
\label{fig: mq_0.4}
(Color online) 
The same plots as in Fig.~\ref{fig: mq_0.1} at $J_{\parallel}=0.4$. 
}
\end{center}
\end{figure}

Figure~\ref{fig: PD} shows the magnetic phase diagram against $J_{\parallel}$ and $H$ at $D=0.05$ and $A^{\rm ion}=0.2$, which is obtained by performing simulated annealing down to $T/J=0.001$. 
There are nine different phases in the phase diagram in addition to the fully-polarized state and the AF-SkX', the latter of which is denoted as the rhombus symbol in Fig.~\ref{fig: PD} [see the caption in Fig.~\ref{fig: PD}]. 
As the AF-SkX' state appears only at two points in the phase diagram, we omit the detailed spin configurations on this phase.
The spin and scalar chirality in real space for nine phases are shown in Figs.~\ref{fig: spin} and \ref{fig: spin2} and the corresponding spin structure factor in momentum space is shown in Figs.~\ref{fig: Sq} and \ref{fig: Sq2}. 
In the left and middle right panels of Figs.~\ref{fig: spin} and \ref{fig: spin2}, the arrows represent $(S_i^x, S_i^y)$, while the colors represent $S_i^z$. 
In the middle left and right panels of  Figs.~\ref{fig: spin} and \ref{fig: spin2}, the color represents $\chi_{\bm{R}}$. 
In Figs.~\ref{fig: Sq} and \ref{fig: Sq2}, the square root of $S_{s\eta}^\alpha(\bm{q})$ for $\alpha=x,y,z$ and $\eta=$A, B is shown. 
We also show the $H$ dependence of the uniform magnetization $M^\alpha_\eta$, $Q_{\nu}$-resolved magnetic moments $\bm{m}_{\eta\bm{Q}_\nu}$, and scalar chirality $\chi^{\rm sc}_\eta$ per layer at $J_{\parallel}=0.1$ in Fig.~\ref{fig: mq_0.1} and $J_{\parallel}=0.4$ in Fig.~\ref{fig: mq_0.4}. 
In Figs.~\ref{fig: mq_0.1} and \ref{fig: mq_0.4}, the data of $\bm{m}_{\eta\bm{Q}_\nu}$ in each ordered state are appropriately sorted for better readability. 

For $H=0$, the single-$Q$ (1$Q$) I state appears for $J_{\parallel} \geq 0$. 
As shown in the real-space spin and chirality configurations in each layer in Fig.~\ref{fig: spin}(a), the spin configurations in both layers are characterized by the single-$Q$ spiral state along the $\bm{Q}_3$ direction. 
Due to the antiferromagnetic interlayer exchange interaction, the $xy$ spins in layers A and B align antiparallel to each other. 
The spiral plane is fixed by $D$ and $A^{\rm ion}$: 
The former tends to favor the out-of-plane cycloidal spiral along the $\bm{Q}_\nu$ direction, while the latter tends to favor the in-plane cycloidal spiral. 
As $D$ is smaller than $A^{\rm ion}$ in the present model parameter, the $m^z_{\eta\bm{Q}_\nu}$ is much small than $m^{x}_{\eta\bm{Q}_\nu}$ and $m^{y}_{\eta\bm{Q}_\nu}$, as shown in Fig.~\ref{fig: Sq}(a).  
In other words, the spiral plane almost lies in the $xy$ plane at zero field. 
The helicity of the out-of-plane cycloidal spiral is opposite for two layers due to the opposite sign of $D$, while that of the in-plane one is the same. 

Let us discuss a phase sequence against the in-plane magnetic field. 
We first examine the case for the small interlayer exchange interaction $J_{\parallel}=0.1$. 
When the in-plane magnetic field is applied along the $x$ direction, the spiral states with $\bm{Q}_2$ and $\bm{Q}_3$ show smaller energy than that with $\bm{Q}_1$. 
This is because the magnetic field tends to favor the spiral, whose plane is perpendicular to the field direction. 
Indeed, $m^z_{\eta\bm{Q}_\nu}$ is slightly enhanced owing to the spin flop against the field direction in Fig.~\ref{fig: mq_0.1}(d). 
In addition to the $x$ component of the uniform magnetization $M^x_{\rm A}=M^x_{\rm B}$ in Fig.~\ref{fig: mq_0.1}(a), the $y$ and $z$ components of the magnetization, i.e., $M^y_{\eta}$ and $M^z_{\eta}$, are induced in both layers; $M^y_{\eta}$ is induced to have the uniform component as $M^y_{\rm A}=M^y_{\rm B}$, while $M^z_{\eta}$ is induced to have the staggered component as $M^z_{\rm A}=-M^z_{\rm B}$, as shown in Fig.~\ref{fig: mq_0.1}(b). 
Meanwhile, the 1$Q$ I state does not have a uniform scalar chirality $\chi^{\rm sc}_{\eta}$ in both zero and finite fields, as shown in Figs.~\ref{fig: spin}(a) and \ref{fig: mq_0.1}(e). 
 
With an increase of $H$, the 1$Q$ I state turns into the Ferri-SkX I state through the first-order phase transition, whose spin and chirality configurations in real space per layer are shown in Fig.~\ref{fig: spin}(b). 
Although the spin configurations in both layers are characterized by the triple-$Q$ modulations, they show layer dependence, as shown in Fig.~\ref{fig: Sq}(b). 
Thus, there are both uniform and staggered components in the magnetization along the $y$ and $z$ directions, as shown in Fig.~\ref{fig: mq_0.1}(b). 
In addition, both layers exhibit nonzero scalar chirality with different magnitudes in Fig.~\ref{fig: mq_0.1}(e). 
Especially, one finds that the spin configuration on layer B corresponds to the SkX when calculating the skyrmion number.  
The skyrmion number in layer B becomes $-1$, while that in layer A becomes zero. 
Thus, this state is regarded as the ferri-type SkX consisting of the skyrmion layer and the topologically-trivial magnetic layer. 
It is noted that the sign of the skyrmion number in the SkX layer is arbitrary, which indicates that the skyrmion and anti-skyrmion are energetically degenerate. 
The similar states consisting of the SkX layer and the other magnetic layer have been realized in the trilayer system~\cite{Hayami_PhysRevB.105.184426} and the threefold-screw symmetric system~\cite{Hayami_PhysRevB.105.224411}. 

When $H$ is further increased, both layers exhibit the SkX spin configuration, as shown in Fig.~\ref{fig: spin}(c). 
The spin structure factor in this state shows the same triple-$Q$ peak structures for layers A and B, as shown in Fig.~\ref{fig: Sq}(c). 
Meanwhile, there are two clear differences in terms of the spin and chirality configurations between the two layers: One is the sign of the $z$-spin component and the other is the sign of the scalar chirality. 
In the case of Fig.~\ref{fig: spin}(c), the skyrmion numbers in layers A and B are $-1$  and $+1$, respectively; the total skyrmion number in the whole system becomes zero. 
Similar to the skyrmion number, the total scalar chirality becomes zero owing to the cancellation between the two layers, $\chi^{\rm sc}_{\rm A}=-\chi^{\rm sc}_{\rm B}$ in Fig.~\ref{fig: mq_0.1}(e). 
Thus, this state well corresponds to the AF-SkX, where the topological spin Hall effect is expected without the topological Hall effect. 
In this sense, this state is qualitatively different from the sublattice-dependent SkXs in previous studies, which have nonzero scalar chirality in the whole system~\cite{Rosales_PhysRevB.92.214439, Diaz_hysRevLett.122.187203, Osorio_PhysRevB.96.024404, LIU201925,gao2020fractional, liu2020theoretical, Tome_PhysRevB.103.L020403, mukherjee2021antiferromagnetic, Mukherjee_PhysRevB.103.134424, Rosales_PhysRevB.105.224402}. 
In addition, the present SkX is also different from the {\it antiferromagnetic} SkX parametrized by the N\`eel order describing the bipartite magnetic lattice~\cite{zhang2016antiferromagnetic, Barker_PhysRevLett.116.147203, Bessarab_PhysRevB.99.140411, Kravchuk_PhysRevB.99.184429,Zarzuela_PhysRevB.100.100408,legrand2020room}.

The appearance of the AF-SkX is due to the synergy among $D$, $A^{\rm ion}$, $J_{\parallel}$, and $H$ in the model in Eq.~(\ref{eq:Ham}). 
The easy-plane single-ion anisotropy $A^{\rm ion}$ and the in-plane magnetic field $H$ play an important role in stabilizing the SkX in each layer. 
In fact, these two ingredients become sources of the SkX when $D=0$ and $J_{\parallel}=0$, although the skyrmion and anti-skyrmion are energetically degenerate in this case ~\cite{Hayami_PhysRevB.103.224418}. 
In other words, the skyrmion number in the whole system takes the values from $+2$ to $-2$ depending on initial spin configurations in the simulations. 
By introducing $D$ and $J_{\parallel}$, such degeneracy is lifted and the relative relationship of the skyrmion number in each layer is determined; one of the layers takes the skyrmion number of $+1$, while the other takes that of $-1$. 

The increase of $H$ in the AF-SkX region leads to the transition to another ferri-type SkX denoted as the Ferri-SkX II in Fig.~\ref{fig: PD}. 
Similar to the Ferri-SkX I, one of the layers is characterized by the SkX with the skyrmion number of $\pm 1$, while the other is characterized by the topologically trivial triple-$Q$ state without the skyrmion number. 
The real-space spin and scalar chirality configurations in Fig.~\ref{fig: spin}(d) and the spin structure factor in Fig.~\ref{fig: Sq}(d) are also similar to those in the Ferri-SkX I. 
The difference between them is found in the quantitative value of the scalar chirality, as shown in Fig.~\ref{fig: mq_0.1}(e). 
In the Ferri-SkX II, the scalar chirality in the trivial triple-$Q$ layer is negligibly small; it takes at most 0.004. 
On the other hand, the trivial triple-$Q$ layer in the Ferri-SkX I exhibits relatively large scalar chirality from $0.02$ to $0.1$. 

The Ferri-SkX II state is replaced by the 3$Q$ I state with jumps of $M^{\alpha}_\eta$, $m^{\alpha}_{\eta\bm{Q}_\nu}$, and $\chi^{\rm sc}_\eta$ by increasing $H$, as shown in Figs.~\ref{fig: PD} and \ref{fig: mq_0.1}. 
In this state, the spin configurations in the two layers are the same except for the relative phase of the spin density waves, as shown in Fig.~\ref{fig: spin2}(a); they are mainly characterized by the single-$Q$ spiral modulation along the $\bm{Q}_3$ direction. 
As shown in the spin structure factor in Fig.~\ref{fig: Sq2}(a), the dominant peaks appear at $\bm{Q}_3$ in $S_{s\eta}^{y}(\bm{q})$ and $S_{s\eta}^{z}(\bm{q})$, while the sub-dominant peaks appear at $\bm{Q}_1$ in $S_{s\eta}^{x}(\bm{q})$ and $S_{s\eta}^{z}(\bm{q})$ and $\bm{Q}_2$ in $S_{s\eta}^{y}(\bm{q})$ and $S_{s\eta}^{z}(\bm{q})$. 
Although this triple-$Q$ spin texture accompanies the scalar chirality density waves shown in Fig.~\ref{fig: spin2}(a), it does not have a uniform component of the spin chirality in Fig.~\ref{fig: mq_0.1}(e)

With increasing $H$, the intensities at the sub-dominant $\bm{Q}_\nu$ gradually become small, and the 1$Q$ II state appears. 
In contrast to the 1$Q$ I state, the spin configuration is characterized by the spiral on the $yz$ plane to gain the energy by the in-plane magnetic field, as shown in Figs.~\ref{fig: spin2}(b) and \ref{fig: Sq2}(b). 
This state continuously turns into the fully-polarized state, as shown in Fig.~\ref{fig: mq_0.1}. 

When $J_{\parallel}$ is increased, the Ferri-SkX I, AF-SkX, Ferri-SkX II, and the 3$Q$ I state vanish, which indicates that the small energy scale of $J_{\parallel}$ compared to $J$ is important to stabilize these states. 
For relatively large $J_{\parallel}$, additional three phases appear in the phase diagram in Fig.~\ref{fig: PD}: 3$Q$ II, 1$Q$ III, and 1$Q$ IV states. 
The $H$ dependence of spin- and chirality-related quantities in these phases is shown in the case of $J_{\parallel}=0.4$ in Fig.~\ref{fig: mq_0.4}.

The 3$Q$ II state appears in the intermediate-field region, which is obtained by increasing $H$ in the 1$Q$ I state for $J_{\parallel} \gtrsim 0.32$. 
The phase transition is of first order with jumps of $M^{\alpha}_\eta$, $m^{\alpha}_{\eta\bm{Q}_\nu}$, and $\chi^{\rm sc}_\eta$; see Fig.~\ref{fig: mq_0.4}(a) for example. 
Similar to the Ferri-SkX I and II, this state consists of layers with different spin configurations accompanying the uniform scalar chirality, as shown in Figs.~\ref{fig: spin2}(c) and \ref{fig: mq_0.4}(e). 
However, there are no skyrmion numbers in both layers. 
As shown in the spin structure factor in Fig.~\ref{fig: Sq2}(c), both spin configurations are characterized by the dominant peak at $\bm{Q}_3$ and the sub-dominant peaks at $\bm{Q}_1$ and $\bm{Q}_2$, but their amplitudes are different for different layers, as shown in Figs.~\ref{fig: mq_0.4}(c) and \ref{fig: mq_0.4}(d).

The 3$Q$ II state is replaced by the 1$Q$ III state at $H\simeq 1.05$. 
The spin configuration in this state is characterized by the single-$Q$ spirals along the $\bm{Q}_1$ direction, as shown in Figs.~\ref{fig: spin2}(d) and \ref{fig: Sq2}(d). 
Compared to the spin configurations for layers A and B, the $y$ and $z$ components of the spin align antiparallel, while the $x$ component of the spin aligns parallel. 
This state exhibits a staggered magnetization parallel to the $y$ direction, $M^y_{\rm A}=-M^y_{\rm B}$, as shown in Fig.~\ref{fig: mq_0.4}(b). 

The further increase of $H$ changes the single-$Q$ ordering vector from the $\bm{Q}_1$ to $\bm{Q}_2$ or $\bm{Q}_3$ for $H \gtrsim 1.3$, which indicates the phase transition to the 1$Q$ IV state.  
As shown in the real-space spin configuration in Fig.~\ref{fig: spin2}(e) and the momentum-space spin structure factor in Fig.~\ref{fig: Sq2}(e), this state is similar to the 1$Q$ II state shown in Figs.~\ref{fig: spin2}(b) and \ref{fig: Sq2}(b). 
Meanwhile, the behavior of $m^z_{\eta\bm{Q}_{\nu}}$ is different from each other: $m^z_{{\rm A}\bm{Q}_{3}} \neq m^z_{{\rm B}\bm{Q}_{3}}$ for the 1$Q$ IV state, while $m^z_{{\rm A}\bm{Q}_{3}} = m^z_{{\rm B}\bm{Q}_{3}}$ for the 1$Q$ II state, as shown in Figs.~\ref{fig: mq_0.1}(d) and \ref{fig: mq_0.4}(d). 
In addition, the 1$Q$ IV state has the $y$-spin component of the staggered magnetization in Fig.~\ref{fig: mq_0.4}(b), while the 1$Q$ II state has that of the uniform magnetization in Fig.~\ref{fig: mq_0.1}(b). 
When $H$ is further increased, this state continuously turns into the 1$Q$ II state. 
It is noted that there is no almost $z$-spin modulation at $\bm{Q}_3$ in the 1$Q$ II state for large $J_{\parallel}$; this spin configuration is almost identified as the single-$Q$ fan state. 
Finally, the fully-polarized state appears for $H\gtrsim 2.85$.

\section{Summary}
\label{sec: Summary}

We have investigated the possibility of generating the AF-SkX without the uniform scalar chirality (skyrmion number) in synthetic layered antiferromagnets. 
By focusing on the centrosymmetric bilayer structure with the local DM interaction, we find that the AF-SkX can be stabilized in the external magnetic field. 
In addition, we obtain the instability toward the ferri-type SkX and other multiple-$Q$ states depending on the model parameters. 
Our study indicates that the rich topological spin textures emerge from the competing exchange interactions including magnetic anisotropy. 
The necessary ingredients for the AF-SkX are (1) the staggered DM interaction, (2) the easy-plane single-ion anisotropy, (3) the interlayer exchange interaction, and (4) the in-plane magnetic field.

Based on the above necessary ingredients, let us discuss an important experimental situation to realize the AF-SkX. 
One is the lattice structure with the staggered DM interaction. 
As the sublattice-dependent feature of the DM interaction is essential, not only the bilayer systems but also the bulk systems can be a candidate. 
For example, the hexagonal system, where the space group belongs to $P6/mmm$ and the magnetic sites lie at the $2e$ site, is one of the candidate systems to possess the staggered DM interaction. 
In addition, a large distance between the different sublattices would be desired, since the small interlay exchange interaction tends to stabilize the AF-SkX. 
For magnetic ions, heavier elements would be better to make the magnitudes of the easy-plane single-ion anisotropy and the DM interaction larger. 
Once such a situation is satisfied, one can obtain the AF-SkX when an external magnetic field is applied along the in-plane direction.

\begin{acknowledgments}
This research was supported by JSPS KAKENHI Grants Numbers JP21H01037, JP22H04468, JP22H00101, JP22H01183, and by JST PRESTO (JPMJPR20L8). 
Parts of the numerical calculations were performed in the supercomputing systems in ISSP, the University of Tokyo.
\end{acknowledgments}

\bibliographystyle{JPSJ}
\bibliography{ref}

\begin{thebibliography}{100}
\bibitem{comment_contour2}
The color in each pixel is plotted by taking the average of $S_i^z$ for four
  neighboring spins located at $\bm{r}_i$, $\bm{r}_i+(1,0,0)$,
  $\bm{r}_i+(1/2,\sqrt{3}/2,0)$, and $\bm{r}_i+(3/2,\sqrt{3}/2,0)$ for better
  visualization, where $\bm{r}_1$ represent the position vector at site $i$.
\end{thebibliography}

\end{document}